\documentclass[oneside,10pt,a4paper]{article}      
\usepackage{amsmath}
\usepackage{amssymb}
\usepackage{graphicx}

\usepackage{lineno}

\usepackage[margin=16mm]{geometry}

\title{\bf Concepts and Criteria for Blind Quantum Source Separation}
 
\author{Alain Deville$^1$, Yannick Deville$^2$ \\
\vspace{1mm}\\
$^1$Aix-Marseille Universit\'e, CNRS, IM2NP UMR 7334, Avenue \\Escadrille Normandie Niemen, 13397, Marseille, France \\ $^2$ Universit\'e de Toulouse,
CNRS-OMP IRAP UMR 5277, 14 Avenue\\Edouard Belin, 31400, Toulouse, France \\
email addresses: alain.deville@univ-amu.fr and yannick.deville@irap.omp.eu
} 

\date{}      

\begin{document}             

\maketitle                   

\begin{abstract}
\normalsize Blind Source Separation (BSS) is an active domain of Classical Information Processing. The development of Quantum Information Processing has made possible the appearance of  Blind Quantum Source Separation (BQSS). This article discusses some consequences of the existence of the entanglement phenomenon, and of the probabilistic aspect of quantum measurements, upon BQSS solutions. It focuses on a pair of spins initially separately prepared in a pure state, and 
then with an undesired coupling between these spins. An unentanglement criterion is established for the state of an arbitrary qubit pair, expressed first with probability amplitudes and secondly with probabilities. It is stressed that the concept of statistical independence of the sources, widely used in classical BSS, should be used with care in BQSS, and possibly replaced by some disentanglement principle. It is shown that the coefficients of the development of any qubit pair pure state, over the states of the standard basis, can be expressed with the probabilities of results in the measurements of well-chosen spin components. 
\end{abstract}
\section{Introduction}\label{SectionIntroduction} 
    The problem of Source Separation (SS), with its so-called Blind version, was introduced around 1985, in the context of Classical Information Processing, and has favored the introduction of concepts and the development of specific methods since then \cite{Comon2010}. Typically, at first, a set of users (the Writer) presents a set of simultaneous signals (called input signals, or sources) at the input of a multi-user communication system hereafter called the Mixer. The sources, constrained to possess some general properties, e.g. mutual statistical independence, are mixed (in the BSS sense, i.e. combined) in the Mixer, often specified through a model, e.g. the simplest one, the linear memoryless model  (cf. Ch. 11 from Ref. \cite{Deville2011}). Another set of users (the Reader) receives the signals arriving at the Mixer output. The Writer knows the sources, but the Reader does not know them, and cannot access the inputs of the Mixer. That Mixer uses one or several parameter values, unknown to the Reader, who only knows some general properties of that Mixer.
The Reader's final task is the restoration of the sources (possibly up to some so-called acceptable indeterminacies) from the signals at the Mixer output, during the inversion phase. An intermediate task is the determination of the unknown parameters of the Mixer, or of its inverse. Before receiving the signals to be separated at the Mixer output, derived from the sources sent by the Writer, the Reader therefore enters an $ ``$adaptation phase", during which he knows that the Writer is sending one (or possibly a limited number of) signal(s) submitted to some definite, and known by the Reader, constraints. The particular signal sent is not known by the Reader (blind separation problem), who knows the class of the input signal(s) and the signal(s) at the Mixer output in the adaptation phase, and, of course, the mixed signals to be separated in the inversion phase.

As any classical phenomenon, conventional Source Separation may be seen as the limit of a quantum phenomenon. In 2007, we began extending SS into the quantum context \cite{Deville2007}, and we have been building up solutions since then (see e.g. Refs. \cite{Deville2012} - \cite{Deville2016Banff}). The aim of this article is to clarify concepts and justify properties already used in our previous papers upon BQSS, a task postponed up to now, possibly deriving new results which could be of use in the BQSS context. In the following sections, some aspects of our previous papers are occasionally mentioned, but the building of any specific BQSS solution is outside the scope of this work. It is hoped that some results established hereafter could also be used in other contexts than BQSS.
Up to now, we have considered the following situation, and we keep to it in this article: at an initial time $t_0$, the Writer prepares two distinguishable qubits numbered 1 and 2, each in a given pure state
\begin{equation}\label{psi1ou2àt0} 
\mid \psi_i(t_0)>=\alpha_i \mid 0>+ \beta_i \mid 1>, \qquad  i=1,2
\end{equation}
where $\mid 0>$ and $\mid 1>$ are orthonormal states, and  $\mid \psi_i(t_0)>$ is normed. These initial quantum states carry information, an idea contained in the expression $ ``$quantum sources". The initial state of the qubit pair is then
\begin{equation}
\mid \Psi(t_0)>=\mid \psi_1(t_0)> \otimes \mid \psi_2(t_0)>.
\end{equation}
The time between $t_0$ (writing) and $t_1$ (reading) is supposed short enough for the qubit pair to be treated as isolated, a choice already made by Feynman \cite{Feynman1985,Feynman1996} in the context of the quantum computer.
At any time $t$ between $t_0$ and $t_1$, the state of the qubit pair may then be described by a ket $\mid \Psi(t)>$. In the Schr\"{o}dinger picture, this time evolution of the pair is described by a time-dependent unitary operator $U(t_0,t_1)$. It is assumed that an undesired coupling exists between these qubits. Because of this undesired coupling, as time goes on the state of the pair generally becomes entangled. Coupling is then interpreted as a mixing (again in the SS sense), realized by an abstract Mixer depending upon one or several parameter values, unknown to the Reader, who only knows some general properties of that Mixer. It is said that the input of the Mixer receives state $\mid \Psi(t_0)>$, and that its output provides state $\mid \Psi(t)>$. It should be well appreciated that inverting $U(t_0,t_1)$ in order to get $\mid \Psi(t_0)>$ from $\mid \Psi(t_1)>$ is not that easy, because $U(t_0,t_1)$ is unknown (blind SS).

When developing solutions to the BQSS problem, one may try and import concepts and methods from the classical to the quantum SS context. However, the presence of entanglement should be clearly identified and the consequences of its existence should not be underestimated. Besides, the concepts of quantum sources and of their statistical independence deserve some discussion, and consequences of the probabilistic aspect of the results of measurements in the quantum domain must be drawn. In Section \ref{SectionUnentanglementCriterion}, it is first explained why quantum tomography is unable to solve the present BQSS problem, and secondly why the Schmidt criterion is ill-suited for following the degree of entanglement of $\mid \Psi(t_1)>$ during the adaptation phase. The Peres-Horodecki criterion \cite{Peres1996,Horodecki1996} is valid for separable mixed (in the quantum sense) states of bipartite systems, and not specifically for unentangled pure states. A better suited unentanglement criterion is therefore established. In Section \ref{SectionModelStatistIndependence}, a model situation, for a single spin and then for a pair of spins, in inhomogeneous magnetic fields with random directions, allows us to speak of random and possibly independent variables, in that quantum context, and to speak of a random quantum state. In Section \ref{SectionCriterionWithProbabilities}, we discuss questions related to the probabilities of the possible results obtained in measurements of spin components, in the context of spins 1/2 as qubits. We first present their use when the Reader makes measurements at the Mixer output in order to restore the sources. These measurements establish a link between the output of the Mixer and the classical world. It is stressed that while the macroscopic support of the results of measurements has a classical behavior, the probabilities of these results obey quantum laws. We then establish an unentanglement criterion using probabilities, equivalent to the one established in Section  \ref{SectionUnentanglementCriterion} for the probability amplitudes $c_i$. It is shown that the $c_i$ coefficients can be expressed as functions of the probabilities of results in the measurements of well-chosen spin components. In Section \ref{SectionDisentanglementHeisenberg}, we derive the expression of the above unentanglement criterion for all possible source states, at the output of the so-called separating system, with respect to the parameters of both the cylindrical Heisenberg coupling and that separating system.
\section{An unentanglement criterion for a qubit pair}\label{SectionUnentanglementCriterion} 
A superficial look may suggest that it is possible to restore the initial product state through State or Process Tomography (ST, PT). ST aims at determining a quantum state if a lot of copies of that state are available \cite{Nielsen2000}. But in BQSS the Reader is unable to access the input of the Mixer, and ST is therefore obviously presently strictly useless. PT would presently consist of placing (preparing) successive well-defined and known quantum states at the input of the Mixer, thus operating in the non-blind mode (cf. Ref. \cite {Deville2011}, page 202) and observing the corresponding signals at its output. 
But, in the BQSS problem, the Reader is strictly unable to operate that way, as he is unable to ask the Writer to prepare him the quite specific input states asked for by PT. Therefore quantum tomography is unable to solve the BQSS problem, which needs dedicated methods (for more details see Ref. \cite {Deville2012}).\\
Up to now, in the BQSS problem, we developed two main approaches for both determination of the unknown parameter(s) of the mixing or separating system and source separation. In the first approach \cite{Deville2007,Deville2012,Deville2014Springer}, the Reader measures observables, using the signals at the Mixer output. The results, and properties associated with them, e.g. the probabilities of their occurences, are kept upon a macroscopic device, e.g. the memory of a classical computer, and then used in a separating system. Since this macroscopic device and the separating system have a classical behavior, we called this processing aimed at restoring the sources $ ``$classical-processing BQSS". In the second, quite different, and more recently introduced approach  \cite{Deville2013,Deville2014Florence}, the quantum state at the Mixer output is sent to the input of a quantum-processing system, the inverting block of the separating system. This block is so designed that its output provides a quantum pure state equal to $\mid \Psi (t_{0})>$ (possibly up to some acceptable indeterminacies), after the adaptation phase.

From now on, the state spaces of two arbitrary qubits, again called qubits 1 and 2, are denoted as $\mathcal{E}_{1}$ and $\mathcal{E}_{2}$ respectively. The possible (pure) states of the pair are the kets in $\mathcal{E}_{1}\otimes $ $\mathcal{E}_{2}.$ We assume that the qubits are physically realized with spins 1/2, which e.g. allows us to speak of the spin component $s_{1z}$ or $s_{2z},$ but many results to be established keep true without this assumption. We introduce the orthonormal basis $\mathcal{B}_{+}$, \{$\mid ++>,$ $\mid +->,$ $\mid
-+>,\mid -->$\}, where e.g. $\mid +->$ means $\mid 1+>\otimes \mid 2->$ and $\mid i,+>$, $\mid i,->$ are normed eigenkets of the $s_{iz}$ component of (reduced) spin $\overrightarrow{s_{i}}$ (with $i$ = 1, 2), for the eigenvalues +1/2 and -1/2 respectively. Any pure pair state, entangled or not, may be expanded in $\mathcal{B}_{+}$ as
\begin{equation}\label{definition ci(t)}
\mid \Psi>=c_1\mid ++>+c_2\mid +->+c_3\mid-+>+c_4\mid -->,
\end{equation}
where the complex coefficients $c_j$ ($j$ = 1 to 4) respect $\sum_j$ $\mid c_j\mid ^2$ $=1$. If a pure or mixed state of a bipartite system $S_{12}$ (parts $S_{1}$ and $S_{2}$) is described by a density operator $\rho$, the corresponding reduced traces $\rho _{1}=$ $Tr_{2}\rho $ and $\rho_{2} $ $=Tr_{1}\rho $ have all the mathematical properties of a density operator \cite{CohenTannoudji1973}. And if $S_{12}$ is in a pure state, $\rho _{1}$ and $\rho _{2}$ have the same eigenvalues \cite{Buchleitner2009}. That pure state is unentangled if and only if its Schmidt number $N_{S}$ (the number of non-zero eigenvalues of $\rho _{1}$ and $\rho _{2}$) is equal to 1 \cite{Buchleitner2009}. We are particularly interested in the case when $\mid \Psi>$ is the state found at the output of the inverting block. Then, any pure state may be expanded in the standard basis $\mathcal{B+}$ as in Eq. (3), where the values of the $c_{i}$ coefficients are affected by both the coupling between the qubits and, during the adaptation phase, by the adaptation procedure. This adaptation phase typically consists of an iterative numerical algorithm which aims at optimizing a continuous-valued function, traditionally called the ``cost function''. For any given values of the adjustable parameters of the inverting block, the cost function measures a kind of $ ``$distance" between $\mid \Psi>$ at the output of the inverting block  and an unentangled pure state. The Schmidt unentanglement criterion cannot be used in our problem, because the considered state remains (at least slightly) entangled throughout the adaptation procedure, and the Schmidt number thus remains higher than one. The Schmidt criterion provides a binary-valued unentanglement detector, with a Schmidt number equal to one or not and, if taking into account all possible integer values of $N_{S}$ beyond unentanglement detection, the Schmidt criterion provides a discrete-valued quantity. What we eventually need instead is a quantitative, continuous-valued, measure of that $ ``$distance" of the considered state with respect to unentanglement, in order to keep the adjustable parameter values of the inverting block yielding the state which is the closest to unentanglement. Moreover, even if the Schmidt approach could be modified to this end, it would yield high computational complexity, as it would require one to diagonalize $\rho_{1}$ or $\rho_{2}$ for each of the quite numerous steps of the iterative adaptation algorithm. We avoid these issues as follows. Since the qubit pair is in a pure state, its partial traces  $\rho_{1}$ and $\rho_{2}$ satisfy
\begin{equation}
\mathrm{Tr}\rho _{1}^{2}=\mathrm{Tr}\rho _{2}^{2}\leq 1,  \label{CritereConnu}
\end{equation}%
and the common value for $\mathrm{Tr}\rho _{1}^{2}$ and $\mathrm{Tr}\rho _{2}^{2}$ is 1 if and only if the pure state is unentangled (cf. Ref. \cite {Buchleitner2009}). One could think of using $\mathrm{Tr}\rho _{1}^{2}-1$ as a cost function. But $\mathrm{Tr}\rho _{1}^{2}$ depends upon the $c_{i}$, which suggests one to try and establish an unentanglement criterion using the $c_{i}$ explicitly. To this end, we consider state  $|\Psi\rangle$ defined through Eq. (\ref{definition ci(t)}). When it is assumed that $|\Psi\rangle$ is unentangled, i.e. that it can be written as
\begin{equation} \label{PsiEtatProduit} 
|\Psi \rangle=(a |+\rangle + b |-\rangle) \otimes (c |+\rangle + d |-\rangle),
\end{equation}
then, in Eq. (\ref{definition ci(t)}), $c_{1}$ = $ac$, $c_{2}$ = $ad$, $c_{3}$ = $bc$, $c_{4}$ = $bd$, so $c_{1}c_{4}$ and $c_{2}c_{3}$ are both equal to $abcd$: 
\begin{equation}\label{c1c4=c2c3} 
c_{1}c_{4} = c_{2}c_{3}.
\end{equation}
Conversely, when it is assumed that Eq. (\ref{c1c4=c2c3}) is satisfied, if $c_{1}\neq 0$ then $|\Psi \rangle$ may be written as
\begin{equation}\label{c1enFacteurs} 
|\Psi \rangle=c_{1}( |+\rangle + \frac{c_{3}}{c_{1}} |-\rangle) \otimes ( |+\rangle + \frac{c_{2}}{c_{1}}|-\rangle  ),
\end{equation}
which means that $|\Psi \rangle$ is then unentangled. If Eq. (\ref{c1c4=c2c3}) is satisfied and $c_{1}=0$, then $c_{2}=0$ and $c_{3}\neq0$, or $c_{3}=0$ and $c_{2}\neq0$, or $c_{2}$ = $c_{3}$ = 0,
and  in each case $|\Psi \rangle$ is unentangled.
Therefore, if the qubit pair is in a pure state $|\Psi \rangle$ written as in Eq. (\ref{definition ci(t)}), then:
\begin{equation}\label{CritèreNonIntri} 
|\Psi \rangle \;\mathrm{is \:unentangled} \Longleftrightarrow  c_{1}c_{4} = c_{2}c_{3}.
\end{equation}
This unentanglement criterion for a qubit pair pure state was used without justification in Refs. \cite{Deville2013} and \cite{Deville2014Florence}.

In Eq. (\ref{definition ci(t)}), $|\Psi \rangle$ was expanded in the standard basis. It is possible instead to introduce e.g. the normed eigenvectors of $s_{1x}$ and $s_{2x},$ or more generally those of $s_{1u}$ and $s_{2v},$ the components of the spins along respective arbitrary directions $\overrightarrow{u}(\theta _{1E},$ $\varphi _{1E})$ and $\overrightarrow{v}(\theta _{2E},$ $\varphi _{2E})$, defined through their Euler angles. For each component, the possible results are again $\pm 1/2$.\ The possible results for the pair may be symbolically written as $(+u+v),$ $(+u-v),$ $ (-u+v)$ and $(-u-v),$ and the corresponding probabilities as $P_{1uv},$ $P_{2uv},$\ $P_{3uv},$ $P_{4uv.}$ Eq. (\ref{definition ci(t)}) is replaced by
\begin{equation}
|\Psi \rangle=c_{1uv}| +u+v\rangle+c_{2uv}|+u-v\rangle+
c_{3uv}|-u+v \rangle+c_{4uv}|-u-v\rangle.
\end{equation}%
With the same reasoning within the new basis, (\ref{CritèreNonIntri}) is replaced
by%
\begin{equation}
|\Psi \rangle \;\mathrm{is \:unentangled} \Longleftrightarrow c_{1uv}c_{4uv}=c_{2uv}c_{3uv}.  \label{c1uvc4uv=c2uvc3uv}
\end{equation}%
\section{Random quantum sources and their independence} \label{SectionModelStatistIndependence}
As in Sections \ref{SectionIntroduction} and \ref{SectionUnentanglementCriterion} the qubits are supposed to be physically realized with electron or nuclear spins 1/2. Standard Electron Spin and Nuclear Magnetic Resonance (ESR, NMR) use a non-microscopic number of resonant spins, but methods have been proposed for more than twenty years in order to detect a single spin, particularly with Optically Detected Magnetic Resonance (ODMR \cite{Kohler1993,Gruber1997}) or with Magnetic Resonance Force Microscopy (MRFM \cite{Rugar2004}), and more recently at low temperature (0.5 K) with Spin Excitation Spectroscopy \cite{Otte2008}, or even with ESR, in extreme conditions \cite{Bienfait2015}. These approaches are still under development. Here, anticipating upon advances in spintronics, we rather consider a pair of spins, or even a single spin, submitted to a static magnetic field.

When speaking e.g. of a microwave source for satellite television, one speaks of the device emitting the microwave carrier. Similarly, the expression ``laser source"  generally refers to the device creating the coherent radiation. In conventional SS, ``source" is an abbreviation for ``source signal". And in Quantum SS with abstract qubits corresponding to physical spins 1/2, the word ``source'' does not refer to some atomic beam delivering atoms carrying an electron or nuclear magnetic moment, but still means ``source signal", then referring to some information from the quantum states of these qubits.

In conventional SS, an important concept is that of statistical independence of the sources, at the root of the frequent use of ICA \cite{Hyvarinen2001}. In Refs. \cite{Deville2007,Deville2012,Deville2014Springer}, we postulated the existence of statistically independent quantum sources when using the classical-processing SS defined at the beginning of Section \ref{SectionUnentanglementCriterion}. Hereafter, we show that statistical independence may exist in that context. We first recall that quantum mechanics does consider e.g. random \textit{operators,} defined as operators the matrix elements of which are random  quantities (see the random lattice operators $F^{(q)}$ in the quantum description of the motions of nuclear moments in liquids, in the study of the spin-lattice phenomenon, in Ref. \cite{Abragam1961}). As a simple model situation, a magnetic moment $\overrightarrow{\mu }$ associated with a single electron spin 1/2, with  $\overrightarrow{\mu }$ $=-G$ $\overrightarrow{s}$ (isotropic   $\overline{\overline{g}}$ tensor), placed in a Stern-Gerlach device, is now introduced. The static field is $\overrightarrow{B_{0}}$=$B_{0}\overrightarrow{Z}$, with amplitude $B_{0}$. The system of interest consists of this spin and the magnet. Writing the Zeeman Hamiltonian as $h=-\overrightarrow{\mu } \overrightarrow{B}_{0}$ = $G {B}_{0}s_{Z}$ indicates that while the spin is a quantum object, the magnetic field is treated classically. The Writer first prepares the spin in the $|+Z\rangle$  eigenstate of $s_{Z}$ (eigenvalue $+1/2$). The moment is then received by the Reader, supposed to ignore the direction of $\overrightarrow{B_{0}}$, and who chooses  some direction attached to the Laboratory as the quantization direction, called $z$ (unit vector $\overrightarrow{u_{z}}$) and introduces a Laboratory-tied cartesian reference frame $xyz$, used to define $\theta _E$ and $\varphi_E$, the Euler angles of $\overrightarrow{Z}$. Since the field is treated classically, $\theta _E$ and $\varphi_E$ behave as classical variables, while $s_{Z}$ is an operator. The Reader measures $s_{z}$ $=\overrightarrow{s}\overrightarrow{u_{z}}$ (eigenstates: $| +\rangle $ and $| -\rangle$), and is interested in the probability $p_{+z}$ of getting $+1/2$. An elementary calculation indicates that

\begin{equation}
| +Z\rangle=r| +\rangle + \sqrt{1-r^{2}}e^{i\varphi }|-\rangle,
\label{PsiAvecRetPhi}
\end{equation}
with 
\begin{equation}
r=\cos \frac{\theta _{2E}}{2}, \qquad \varphi=\varphi_{E},
\label{Expressionretphi}
\end{equation}
and therefore  $p_{+z}=\cos^{2}{\theta _{E}/2}.$

Once the direction of the magnetic field has been chosen, state $|+Z\rangle$ is then unambiguously defined. If this direction has a deterministic nature, $r$ and $\varphi$ are deterministic variables, and $| +Z\rangle$ may then be called a deterministic quantum state. If $\theta_{E}$ and $\varphi_{E}$, defining the direction of $\overrightarrow{B_{0}}$ chosen by the Writer, obey probabilistic laws, one may consider that the quantum quantities $r$ and $\varphi$, which depend upon the classical Random Variables (RV) $\theta_{E}$ and $\varphi _{E}$, do possess the properties of conventional, i.e., classical, RV. It may e.g. happen that they be uncorrelated, or even independent\ (which happens if $\theta _{E}$ and $\varphi _{E}$ are independent). And if $\theta _{E}$ and $\varphi _{E}$ depend on time in a random way,  $r$ and $\varphi$ are then random time functions. We are not strictly facing the quantum equivalent of a classical situation here. Rather, the stochastic character of the field direction, with classical nature, is reflected in the random behavior of the quantum state expressed through Eq. (\ref{PsiAvecRetPhi}). Therefore, rather than a random operator, we here meet a random quantum state. The concept of a random state, if not the expression, was already used e.g. in the early and canonical books Refs. \cite{Tolman1938,vonNeumann1946}. The probability $p_{+z}$, presently a function of the RV $\theta _{E},$ is itself an RV. This results from both the randomness of the field direction and the standard probabilistic interpretation of quantum theory. Probabilities of results of measurements for a qubit pair were treated as RV, without the present justification, in most of our previous papers, including Refs. \cite {Deville2007,Deville2012,Deville2014Springer}.

If one measures the scalar observable $O$ when the spin is in the state $|\Psi\rangle=$ $\alpha |+\rangle + \beta | -\rangle$ = $\Sigma_{k}f_{k} |\varphi_{k}\rangle$ (where $k$ is associated with $+$ and $-$), had the $f_{k}$ been deterministic the mean value would have been:%
\begin{equation}
\langle \Psi |O| \Psi \rangle=\sum_{k,l}f_{k}^{\ast}f_{l}O_{kl},    \qquad O_{kl}=\langle \varphi_{k} |O| \varphi_{l} \rangle.
\end{equation}
Since the $f_{k}$ are random, one must moreover calculate the statistical mean, denoted as $\overline{\langle \Psi |O| \Psi \rangle}$: 
\begin{equation}\label{OMoyen}
\overline{\langle \Psi |O| \Psi \rangle}=\sum_{k,l} \overline{f_{k}^{\ast}f_{l}}O_{kl}=Tr{\rho O},
\end{equation}
where $\rho$ is the density operator, whose matrix elements in the $( | +\rangle$ , $| -\rangle ) $ basis are $\rho_{l,k}$ = $\overline{{f_{k}^{\ast}f_{l}}}$. Therefore, it is in principle possible to presently introduce a density operator, which is a non-random operator (its matrix elements are not random quantities, but statistical averages). But this does not present any interest, since in the BQSS problem examined up to now the Reader knows that e.g. qubit 1 has been prepared in a pure state, but does not know the values of the  $\rho_{ij}$ coefficients in any basis, and is consequently unable to choose a basis in which $\rho$ would be diagonal. It is simpler to keep speaking of a random pure state.

As a model situation, we now consider two spins 1/2 numbered 1 and 2, each with conditions similar to the previous ones, with fields along directions with respective unit vectors $\overrightarrow{%
Z_{1}}(\theta _{1E},\varphi _{1E})$ and $\overrightarrow{Z_{2}}(\theta
_{2E},\varphi _{2E})$, and each spin initially prepared in the state
\begin{equation}
| \psi _{i}(t_{0})\rangle=r_{i}|i +\rangle + \sqrt{1-r_{i}^{2}}e^{i\varphi_{i}}|i-\rangle,
\qquad i=1,2,  \label{PsiIavecRiEtPhii}
\end{equation}
where $|i +\rangle$ and $|i -\rangle$ are the eigenkets of $s_{iz}$, 
the component of $\overrightarrow{s_{i}}$ along the quantization direction, for the eigenvalues 1/2 and -1/2 respectively. For the same reason, if the field directions are random, $r_{1},$ $\varphi _{1},$ $r_{2}$ and $\varphi _{2}$ have the properties of conventional RV. If ($\theta _{1E},\varphi _{1E}$) and ($\theta
_{2E},\varphi _{2E}$) are mutually statistically independent, the same is
then true for the couples of RV ($r_{1},$ $\varphi _{1}$)$~$and ($r_{2},$ $%
\varphi _{2}$). And if e.g. $\theta _{1E}$ and $\varphi _{1E}$ are
independent, the same is true for $r_1$ and $\varphi _1$ (cf. Eq. (\ref{Expressionretphi})). These properties are of major importance for our quantum-source independent component analysis (QSICA) methods described in Ref. \cite{Deville2014Springer}. We may then say that the initial state of each qubit is random, i.e. that in Eq. (\ref{PsiIavecRiEtPhii}) $r_{i}$\ and $\varphi _{i}$ are RV.\ This discussion shows that when considering the preparation of a pair of qubits each in a pure state, one may assume either a deterministic or a random direction for each magnetic field. In the latter case, the relevant concept is that of random quantum states, rather than that of random quantum operators mentioned earlier in this section.

Keeping our assumption of a pair of qubits each prepared in a pure state, we now consider the second approach for the adaptation and inversion phases (cf. the beginning of Section \ref{SectionUnentanglementCriterion}), with a quantum state $| \Phi\rangle$ present at the output of the inverting block. The presence of $| \Phi\rangle$ and the Reader's final aim, the recovery of the initial pure state, prompts the Reader: 1) to speak of a deterministic or random pure state, rather than to use a density operator, 2) to consider that the first constraint to be respected in BQSS is then the very existence of an unentangled state at the output of this inverting block. If unentanglement has first been achieved, then and only then is it possible to speak of a deterministic or random state for each part of that product state. While entanglement has no classical counterpart, the following point may be noted here: if a bipartite system is in a pure (deterministic) state $|\Phi\rangle,$ to which a density operator $\rho $ $=|\Phi\rangle \langle\Phi|$ corresponds, $|\Phi\rangle$ is unentangled if and only if the partial traces $\rho _{1}$ and $\rho _{2}$ satisfy the equality $\rho =\rho _{1}\otimes \rho _{2}$ \cite{Barnett2009}. This unentanglement condition is reminiscent of the relation $\rho =$  $\rho_{1}$.$\rho_{2}$ between $\rho $, the joint probability density function of independent classical RV $X_{1}$ and $X_{2},$ and $\rho _{1}$ and $\rho _{2},$ the respective marginal probability density functions. Presently, operators replace functions, a tensor product replaces the ordinary product, and this reminiscence reflects the existence of a classical analogue to unentangled states. Condition (\ref{c1c4=c2c3}) for unentanglement was established using spins 1/2, but is valid for any pair of two-level systems. This discussion suggests that, in the BQSS problem, when considering a pair of qubits prepared in a pure state, and moreover using the second approach of Section \ref{SectionUnentanglementCriterion} for adaptation and inversion, instead of  trying to directly import ICA methods into the BQSS context one should focus upon disentanglement at the output of the inverting block, which recently led us to introduce a disentanglement-based separation principle \cite{Deville2013,Deville2014Florence}.
 
In the next section, use will be made of the number of real independent parameters necessary to define an arbitrary normed ket $|\Psi\rangle$ in $\mathcal{E}_{1}\otimes \mathcal{E}_{2},$ written as in Eq. (\ref{definition ci(t)}), and a ket in $\mathcal{E}_{1}\otimes 
\mathcal{E}_{2}$ forced to be unentangled. These numbers are specified hereafter. An arbitrary\ normed ket $|\Psi\rangle$ in $\mathcal{E}_{1}\otimes \mathcal{E}_{2}$ depends upon the four complex quantities $c_{1}$ to $c_{4}$ linked through two relations between real numbers ($\sum_{i}\mid c_{i}\mid ^{2}$ is equal to $1,$ and $| \Psi\rangle $ and $e^{i\varphi }| \Psi \rangle,$ with $\varphi $ an arbitrary real quantity, should be considered identical). An \textit{arbitrary normed }ket $| \Psi \rangle$ in $\mathcal{E}_{1}\otimes \mathcal{E}_{2}$ therefore depends upon \textit{six} real independent parameters. If it is forced to be
unentangled, it has to satisfy the equality $c_{1}c_{4}=c_{2}c_{3}$ between complex quantities.\ An \textit{unentangled} normed ket $%
| \Psi\rangle$ therefore depends upon \textit{four }real parameters. This corresponds to the fact that $| \Psi \rangle$ is then restricted to the form $|\Psi\rangle=$ $| \psi _{1}\rangle\otimes | \psi _{2}\rangle,$ where the normed kets $| \psi _{1}\rangle$ and $|\psi _{2}\rangle,$ describing the state of qubits $1$ and $2$ respectively, each depend upon two real parameters $%
(r_{1},\varphi _{1}),$ $(r_{2},\varphi _{2})$ (cf. Eq. (\ref{PsiIavecRiEtPhii})).
\section{BQSS and probabilities in spin component measurements} \label{SectionCriterionWithProbabilities}
\subsection{Probabilities in measurements, classical versus quantum world}\label{SubsectionProbaClassQuantumWorlds}
In this subsection, we are interested in our first approach as defined in Section \ref{SectionUnentanglementCriterion}, with measurements at the Mixer output. We specifically consider the solutions to BQSS discussed in Refs. \cite{Deville2007,Deville2012,Deville2014Springer}, with two spins 1/2, each prepared in a pure state at $t_0$, then submitted to an undesired Heisenberg cylindrical coupling \cite{Abragam1961,Fazekas1999} (axial component: $J_{z}$, normal component: $J_{xy}$, cf. Eq. (4) and Appendix E of Ref. \cite{Deville2012}, and Ref. \cite{Abragam1970}), and measurements of $s_{1z}$ and $s_{2z}$ at the output of the formal Mixer at $t_1$. The probabilities of obtaining $(+1/2,+1/2)$, $(+1/2,-1/2)$, $(-1/2,+1/2)$ and $(-1/2,-1/2)$ are denoted respectively as $p_{1},$ $p_{2},$ $p_{3}$ and $p_{4}$ (as in Ref. \cite{Deville2012}, while in Ref. \cite{Deville2007} e.g. our present $p_{4}$ was denoted as $p_{2}$). We keep Eq. (\ref{PsiIavecRiEtPhii}) for both qubits, with the choice $\varphi_{1}=0$. One then gets \cite{Deville2012}:
\begin{equation} \label{p1p4}
p_{1}=r_{1}^{2}r_{2}^{2}, \qquad p_{4}=(1-r_{1}^{2})(1-r_{2}^{2}).
\end{equation}%
$p_{2}$ depends upon a mixing parameter $v = $ sgn$(\cos\Delta _{E})\sin\Delta _{E}$, with \cite{Deville2012} $\Delta_{E}=-J_{xy}(t_1-t_0)/\hbar $. This expression for $\Delta_{E}$ may be vizualized as the opposite of the phase rotation $\Delta \phi$ = $\omega (t_1-t_0)$ between states coupled by a Hamiltonian term with energy $J_{xy}$, during the time interval $(t_1-t_0)$, with $\omega$ given by the Planck-Einstein relation $\omega=J_{xy}/\hbar$. Probability $p_{2}$ satisfies
\begin{equation}
p_{2}=r_{1}^{2}(1-r_{2}^{2})(1-v^{2})+(1-r_{1}^{2})r_{2}^{2}v^{2}-2r_{1}r_{2}%
\sqrt{1-r_{1}^{2}}\sqrt{1-r_{2}^{2}}\sqrt{1-v^{2}}v\sin \Delta _{I}
\label{p2}
\end{equation}
and, with our choice for $\varphi_{1},$ $\Delta _{I}=$\ $\varphi _{2}.$

In Eq. (\ref{PsiIavecRiEtPhii}), which describes the initial state of the qubit pair, $r_{1},$ $r_{2},$ $\varphi _{1}$ and $\varphi _{2}$, are used to define probability amplitudes, i.e. quantum quantities. Expressions (\ref{p1p4}) and (\ref{p2}) show that $p_{1},$ $ p_{4}$ and $p_{2}$ depend upon both $r_{1}$ and $r_{2},$ and that $p_{2}$
moreover depends upon $\Delta _{I}$ and therefore the probabilities clearly follow quantum laws. This instance illustrates the distinction to be made between the quantum status of these probabilities and the classical nature of the laws obeyed by the supports which store them.

In  Refs. \cite{Deville2007,Deville2012,Deville2014Springer}, once $r_{1},$ $r_{2}$ and $\Delta _{I}$ were known, the initially prepared qubit states were completely known, and in the context of classical-processing BQSS we called $r_{1},$ $r_{2}$ and $\Delta _{I}$ the sources (cf. Section \ref{SectionModelStatistIndependence}) in order to focus on the quantities used in the SS process.

The concept of RV is often used in a classical context. Since on the contrary probabilities $p_{1}$,\ $p_{4}$ and $p_{2}$ follow quantum laws, treating them as RV does not go without saying. But Eqs. (\ref{p1p4}) and (\ref{p2}) establish that when $r_{1},$ $r_{2}$, $\varphi _{2}$\ are RV (cf.\ Section \ref{SectionModelStatistIndependence}) the same is true for $p_{1},$ $p_{4}$ and $p_{2}$. They also indicate that $p_{1}$,\ $p_{4}$ and $p_{2}$ depend upon both $r_{1}$ and $r_{2},$ and that $p_{2}$ also depends upon $\Delta _{I}.$ When $J_{xy}=0$ (Ising Hamiltonian $-2Js_{1z}s_{2z}$), then $v=0$ and, for the state at the Mixer output, $p_{1}p_{4}=p_{2}p_{3},$ which can be interpreted as follows. The four states defining the $\mathcal{B}_{+}$ basis are then eigenstates of the Hamiltonian, but time evolution introduces phase differences, and it can be verified that the state at the Mixer output is \textit{entangled} (except if, accidentally, $J(t_1-t_0)/
\hbar=k\pi ,$ $k$ being an integer).
However, when measuring $s_{1z}$ and $s_{2z}$, the probability of getting $(1/2,1/2)$ is then time-independent, which is also true for the probabilities of getting $(1/2,-1/2),$ $(-1/2,1/2)$ or $(-1/2,-1/2).$ Therefore both products $p_{1}p_{4}$ and $p_{2}p_{3}$ are time-independent, and since $p_{1}p_{4}$= $p_{2}p_{3}$ at $t_{0},$ because the qubit pair is then in a product state, this equality is 
preserved as time goes on, although the state has become entangled.\\
In the end, these measurements made at the output of the mixer establish a bridge between the classical and the quantum worlds, the results being kept on macroscopic devices with classical behavior while the probabilities of their occurences follow quantum laws.

\subsection{An unentanglement criterion using probabilities\label{SubSectionProbabilitesEtCritereNonInt}}
The unentanglement criterion expressed through Eq. (\ref{c1c4=c2c3}) uses the $c_{i}$ coefficients, i.e. probability amplitudes. However measurements give access to probabilities, not to probability amplitudes, and the question of establishing whether this unentanglement criterion could be formulated with probabilities  (of the results from spin component measurements) therefore seems relevant. State $|\Phi \rangle$ being present at the ouput of the inverting block, and the components $s_{1u}$ and $s_{2u}$ being then measured, we denote the probabilities of obtaining $(1/2,\ 1/2)$, $(1/2,-1/2$), $(-1/2,\ 1/2)$ and $(-1/2,\ -1/2)$ as $P_{1u},$ $P_{2u},$ $P_{3u},$ $P_{4u}$ respectively, and
the corresponding eigenstates of $s_{1u}$.$s_{2u}$ as $|+u,+u\rangle,$ $|+u,-u\rangle,$ $|-u,+u\rangle$ and $|-u,-u\rangle$. If e.g. $s_{1x}$ and $s_{2x}$ are measured, the probabilities are denoted as $P_{ix}$, with $i=1$ to $4.$ 
In Section \ref{SectionModelStatistIndependence} it was said that an unentangled normed ket $|\Psi \rangle$ in $\mathcal{E}_{1}\otimes \mathcal{E}_{2}$ possesses four degrees of freedom. Taking the squared modulus of each member of the equality $c_1c_4=c_2c_3$ leads to
\begin{equation}
P_{1z}P_{4z}=P_{2z}P_{3z}.  \label{P1zP4z=P2zP3z}
\end{equation}%
Then, taking $\overrightarrow{u}$ and $\overrightarrow{v}$ of Section \ref{SectionUnentanglementCriterion}
both along direction $x,$ we know that $c_{1x}c_{4x}=c_{2x}c_{3x}$ for an unentangled state (cf. Eq. (\ref{c1uvc4uv=c2uvc3uv})), and therefore that
\begin{equation}
P_{1x}P_{4x}=P_{2x}P_{3x}.  \label{P1xP4x=P2xP3x}
\end{equation}%
Eq. (\ref{P1zP4z=P2zP3z}) together with (\ref{P1xP4x=P2xP3x})\ is however
weaker than condition $c_{1}c_{4}=c_{2}c_{3},$ as can be tested by considering
the following state:%
\begin{equation}
|\Psi _{i-i11}\rangle=\frac{1}{2}(i| ++\rangle-i| +-\rangle+|-+\rangle+| --\rangle).
\end{equation}%
$| \Psi _{_{i-i11}}\rangle$ is entangled since $c_{1}c_{4}=-$ $c_{2}c_{3}.$ It
can be written
\begin{equation}
| \Psi _{_{i-i11}}\rangle=\frac{1}{2}(| +x,+x\rangle+i|+x,-x\rangle-|
-x,+x\rangle+i| -x,-x\rangle).  \label{Phi1ProjeteSelonXX}
\end{equation}%
\ Eq. (\ref{Phi1ProjeteSelonXX}) shows that the four probabilities $P_{ix}$
attached to $| \Psi _{_{i-i11}}\rangle$ are all equal to $1/4.$ Therefore $| \Psi _{_{i-i11}}\rangle$ satisfies (\ref{P1zP4z=P2zP3z}) and (\ref{P1xP4x=P2xP3x}), while being entangled.

The two qubits being in the state $| \Psi \rangle$ expressed through (\ref{definition ci(t)}), one may decide to treat the three orthogonal directions on the same footing, measuring successively $s_{x}$ for both spins, then, in a new set of preparations/measurements, $s_{y}$ for both spins, and finally $s_{z}$ for both spins. The probabilities of obtaining ($1/2,1/2),$ $\:$($1/2,-1/2$),$\:$ ($-1/2,1/2),$$\:$(-$1/2,-1/2)$    $\:$respectively, when measuring $s_{1k}$ and $s_{2k}$ (with $k$ successively equal to $x,$ $y,$ and $z$), will be denoted as $P_{1k},$ $P_{2k},$ $P_{3k}$ and $P_{4k}.$ For e.g. the entangled state $\mid \Psi _{_{i-i11}}\rangle$, as $P_{1z}P_{4z}=P_{2z}P_{3z}$ and $P_{1x}P_{4x}=P_{2x}P_{3x},$ the hope is that entanglement can be detected thanks to $P_{1y}P_{4y}$ $\neq$ $P_{2y} P_{3y},$ but in fact the four $P_{iy}$ are equal to $1/4$. Therefore measuring the same spin component for both qubits, successively for $x,$ $y$ and $z$, fails to allow us to build up an unentanglement criterion.

However, since two spins are present, there is still the possibility of not systematically measuring the same spin component for both spins. One chooses to measure successively $s_{z}$ for both spins, then $s_{1z}$ and $s_{2x}$ in a new set of preparations/measurements, and finally $s_{1z}$ and $s_{2y}.$ The presence of the $s_{1z}$ measurement in each of these sets corresponds to recognizing that (\ref{definition ci(t)}) uses the standard basis. The probabilities of obtaining ($1/2,1/2),$ ($1/2,-1/2),$($-1/2,1/2), $ (-$1/2,-1/2)$ respectively when measuring $s_{1i}$ and $s_{2j}$ (with $i=z,$ $x,$ or $y,$ and $j=z,$ $x,$ or $y)$ will be denoted as $P_{1ij},$ $P_{2ij},$ $P_{3ij}$ and $P_{4ij}.$ Denoting the $c_{i}$ introduced in Eq. (\ref{definition ci(t)}) as $c_{i}=\rho _{i}e^{i\psi _{i}},$ then from 
Eq. (\ref{c1c4=c2c3}) it is known that $| \Psi \rangle$ is unentangled if and only if
\begin{equation}
\{\rho _{1}\rho _{4}=\rho _{2}\rho _{3} \qquad \mathrm{and} \qquad
\psi _{1}+\psi _{4}=\psi_{2}+\psi _{3} \ \: \mathrm{mod}\: \mathrm{2\pi} \}.
\label{CNSnonIntricExplicite}
\end{equation}

Measuring \{$s_{1z},$ $s_{2z}$\} allows us to know the moduli
$\mid c_{i}\mid ^{2}=\rho _{i}^{2}$ in (\ref{definition ci(t)}), and to express the first equality in Eq. (\ref{CNSnonIntricExplicite}) as%
\begin{equation}
P_{1zz}P_{4zz}=P_{2zz}P_{3zz}.
\end{equation}%
The $P_{kzx}$ and $P_{kzy}$ (with $k=1$ to $4$), when expressed as functions
of the moduli $\rho _{l}$ and angles $\psi _{m},$ depend upon trigonometric
functions of the $\psi _{m}$ angles.\ For instance, for any state $|\Psi\rangle$ 
entangled or not
\begin{equation}
2P_{1zx}=(\rho _{1}^{2}+\rho _{2}^{2})+2\rho _{1}\rho _{2}\cos (\psi
_{1}-\psi _{2}).
\end{equation}%
When expressing unentanglement through probabilities, one\ then has to try
and respect both $\cos \alpha $ $=$ $\cos \beta $ and $\sin \alpha $ $=$ $%
\sin \beta $ with $\alpha $ and $\beta ~$\ values compatible with the
equality $\psi _{1}+\psi _{4}=\psi _{2}+\psi _{3},$ rather than to respect
the equality $\psi _{1}+\psi _{4}=\psi _{2}+\psi _{3}$ (mod $2\pi $) itself. 
If it is first known that simultaneously $P_{1zz}P_{4zz}=P_{2zz}P_{3zz}$ and $P_{1zx}P_{4zx}=P_{2zx}P_{3zx}$\ are true, then
one immediately deduces that $\cos (\psi _{1}-\psi _{2})=\cos (\psi
_{3}-\psi _{4})$. And if $P_{1zy}P_{4zy}=P_{2zy}P_{3zy}$ replaces the second 
equality, one deduces that $\sin
(\psi _{1}-\psi _{2})=\sin (\psi _{3}-\psi _{4})$. Therefore, when the three
equalities between probability products are satisfied, then $\rho _{1}\rho
_{4}=\rho _{2}\rho _{3}$ and $\psi _{1}+\psi _{4}=\psi _{2}+\psi _{3}$ (mod $%
2\pi $). Conversely, if $| \psi\rangle$ is unentangled, then Eq. (\ref{c1uvc4uv=c2uvc3uv})
implies that $P_{1zj}P_{4zj}=P_{2zj}P_{3zj}$, with $j={z, x, y}$ respectively. 
Finally,%
\begin{equation}
c_{1}c_{4}=c_{2}c_{3} \Longleftrightarrow  \{P_{1zj}P_{4zj}=P_{2zj}P_{3zj}, 
\qquad\mathrm{with} \quad j=x,\:y,\: z \}.
\label{CritereNonIntricAvecProba}
\end{equation}
The equivalence therefore is between a single relation between probability amplitudes
and\ a triplet of relations between probabilities. This criterion, although established 
in the context of BQSS, has the same general validity as Eq. (\ref{c1c4=c2c3}).

Use of criterion (\ref{CritereNonIntricAvecProba}) necessitates successive
measurements first \ of $s_{1z}$ and $s_{2z}$, then (after new
preparations) of $s_{1z}$ and $s_{2x},$ and finally (again after new
preparations) of $s_{1z}$ and $s_{2y}$, in order to successively estimate
first the $P_{izz}$ probabilities, then the $P_{izx}$ and finally the $%
P_{izy}$. One must measure $s_{1z}$ each time, because 1) getting e.g. ($%
+1/2 $, $-1/2$) when measuring $s_{1z}$ and $s_{2z}$ is an event to be
distinguished from the one realized when measuring $s_{1z}$ and $s_{2x}$ and
getting ($+1/2$, $-1/2$), 2) results of\ measurements of $s_{1z}$ and $%
s_{2x} $ are independent only if $|\Psi\rangle$ is unentangled, which
precisely can't be assumed when Eq. (\ref{CritereNonIntricAvecProba}) is to be
used.

The two distinguishable spins were made to play different roles in
the process which led to Eq. (\ref{CritereNonIntricAvecProba}) (systematic measurement
of $s_{1z}$). But this dissymmetry is only apparent, as Eq. (\ref{CritereNonIntricAvecProba}) 
can be replaced by a version obtained by exchanging the spin numbers. Next subsection makes
a symmetrical use of measurements of spin components, allowing one to get the
\textit{values} of both the $\rho _{i}$ moduli and the $\psi _{i}$ angles for the
$c_{i}$ coefficients in Eq. (\ref{definition ci(t)}).
\subsection{Knowing 2-qubit pure states from s$_{ij}$ measurements}\label{ProbaEtAccesEtatpur}
\hspace{0.5cm}If a qubit pair physically realized with spins 1/2 is known to be in an arbitrary pure state described by $|\Psi \rangle$ written as in Eq. (\ref{definition ci(t)}), with $c_{i}=\rho _{i}e^{i\psi _{i}}$ and $i=1$ to $4,$ then in order to know $| \Psi \rangle$ one should know three moduli $\rho _{i}$ and three angles $\psi _{i}$. Accessing these six real quantities is more demanding than testing $|\Psi \rangle$ unentanglement, since once these quantities are known, it is always possible to know whether $|\Psi \rangle$ is unentangled, by testing whether both equalities $\rho _{1}\rho _{4}=\rho _{2}\rho _{3}$ and $\psi _{1}+\psi_{4}$ $=\psi _{2}+\psi _{3}$ are satisfied. On the contrary, when one focuses
upon entanglement, these two equalities may be found to be satisfied, while the values of the $\rho _{i}$ and $\psi _{i}$ are unknown. In the previous subsection, an unentanglement criterion using only probabilities in the measurements of the $s_{ij}$ components, equivalent to the $c_{1}c_{4}=c_{2}c_{3}$ criterion, was given.\ Its existence suggests the
following question: is it possible to access these six real quantities using only probabilities of results in the measurements of the spin components? We are going to show that the answer is yes. It is already known that measurements of both $s_{1z}$ and $s_{2z}$ give access to the moduli $\rho_{i},$ through the probabilities $P_{izz}$ introduced in Section \ref{SubSectionProbabilitesEtCritereNonInt}. One is left with e.g. determining the three angle differences ($\psi _{1}-\psi _{3})$, ($\psi _{2}-\psi _{3})$ and ($\psi _{4}-\psi _{3})$ from well-chosen probabilities. We first consider measurements of $s_{1z}$ and $s_{2i},$ with $i=x$ or $y,$ as in Subsection \ref{SubSectionProbabilitesEtCritereNonInt}. When measuring $s_{1z}$ and $ s_{2x},$ the probabilities of getting $(1/2,$ $1/2$) and $(-1/2,$ $1/2$) are respectively%
\begin{equation}
P_{1zx}=\frac{1}{2}\mid c_{1}+c_{2}\mid ^{2}, \qquad
P_{3zx}=\frac{1}{2} \mid c_{3}+c_{4}\mid ^{2},  \label{P1zxEtP3zx}
\end{equation}%
which leads to
\begin{equation}
\cos (\psi _{1}-\psi _{2})=\frac{2P_{1zx}-P_{1zz}-P_{2zz}}{2\sqrt{P_{1zz}P_{2zz}}},\;
\cos (\psi _{3}-\psi _{4})=\frac{%
2P_{3zx}-P_{3zz}-P_{4zz}}{2\sqrt{P_{3zz}P_{4zz}}}.  \label{CosPhi1-Phi2Et3-4}
\end{equation}
Similarly, when measuring $s_{1z}$ and $s_{2y},$ the probabilities of getting $(1/2,$ $1/2$) and $(-1/2,$ $1/2$) are respectively
\begin{equation}
P_{1zy}=\frac{1}{2}\mid c_{1}-ic_{2}\mid ^{2}, \qquad
P_{3zy}=\frac{1}{2}\mid c_{3}-ic_{4}\mid ^{2},
\end{equation}
which leads to
\begin{equation}
\sin (\psi _{1}-\psi _{2})=-\frac{2P_{1zy}-P_{1zz}-P_{2zz}}{2\sqrt{P_{1zz}P_{2zz}}},\;
\sin (\psi _{3}-\psi _{4})=-\frac{2P_{3zy}-P_{3zz}-P_{4zz}}{2\sqrt{P_{3zz}P_{4zz}}}.
\label{SinPhi1-Phi2Et3-4}
\end{equation}
Expressions (\ref{CosPhi1-Phi2Et3-4}) and (\ref{SinPhi1-Phi2Et3-4}) allow us
to know both $(\psi _{1}-\psi _{2})$ and $(\psi _{3}-\psi _{4})$ (mod $2\pi $%
)$.$

Now, exchanging the roles of spin $1$ and spin $2$, we successively measure \{$s_{1x},$ $s_{2z}\}$ and (after new preparations) \{$s_{1y},s_{2z}\}.$ The probabilities of getting $(1/2,$ $1/2$) in these measurements are respectively
\begin{equation}
P_{1xz}=\frac{1}{2}\mid c_{1}+c_{3}\mid ^{2}, \qquad
P_{1yz}=\frac{1}{2}\mid
c_{1}-ic_{3}\mid ^{2},
\end{equation}%
which lead to
\begin{equation}
\cos (\psi _{1}-\psi _{3})=\frac{2P_{1xz}-P_{1zz}-P_{3zz}}{2\sqrt{P_{1zz}P_{3zz}}},\;
\sin (\psi _{1}-\psi _{3})=-\frac{2P_{1yz}-P_{1zz}-P_{3zz}}{2\sqrt{P_{1zz}P_{3zz}}}.
\label{CosSinPhi1-phi3}
\end{equation}
$(\psi _{1}-\psi _{3})$ is therefore known (mod\ 2$\pi )$.

If one wants to identify not the state at the Mixer input but a pure state at the Inverter output, State Tomography (ST) may in principle be used. But it is far simpler to make measurements for the five \{$s_{1i}$, $s_{2j}$\}\ pairs just considered and to access the corresponding probabilities, than to use ST. The reason is that ST claims to be valid for any quantum state, and therefore does not take advantage of the fact that the qubit pair is presently known to be in a pure state. The dimension of the state space of the qubit pair being four, then for ST one has to introduce sixteen operators, namely the Identity, the six operators $s_{1i}$ and $s_{2j}$ (with $i=x,$ $~y,$ $z$,~and $j=x,$ $~y,$ $z$), and the nine products $s_{1i}s_{2j}$ \cite{Nielsen2000}. One should determine experimentally fifteen mean values, giving access to fifteen independent real values together defining the density operator describing the qubit pair state (three diagonal real elements, and six non-diagonal complex elements).

The simpler state estimation procedure proposed in this section therefore opens the way to new classes of BQSS methods, that we just started to explore in Ref. \cite{Deville2015}, then applying this procedure to the Mixer output.
\section{Disentanglement and cylindrical-symmetry Heisenberg coupling}\label{SectionDisentanglementHeisenberg}
In Subsection \ref{SubsectionProbaClassQuantumWorlds}, we considered measurements made at the Mixer output. We now come to the method for BQSS used e.g. in Ref. \cite{Deville2013}, with classical processing in the adapting block of the separating system, using the notations of Ref. \cite{Deville2013}. $| \Psi (t_{0})\rangle$, the initial product state of the qubit pair, is given by Eq. (\ref{definition ci(t)}), with the values of the coefficients $c_i$\ (in the $\mathcal{B}_{+}$ basis) taken at $t_{0}$ and denoted as $c_i(t_0)$. These components form the source vector
\begin{equation} \label{C+(tzero)}
C_{+}(t_{0})=[c_{1}(t_{0}),c_{2}(t_{0}),c_{3}(t_{0}),c_{4}(t_{0})]^{T},\qquad T: \mathrm{transpose}.
\end{equation}
Similarly, the state at the Mixer output at time $t$, here denoted as $|\Psi (t)>$, is given by Eq. (\ref{definition ci(t)}), with the values of the coefficients $c_i$ (in the $\mathcal{B}_{+}$ basis) taken at $t$ and denoted as $c_i(t)$. The coupling-induced transition from state $| \Psi (t_{0})\rangle$ to $| \Psi (t)\rangle$ is interpreted as the transformation induced by the Mixer, leading to the appearance of $| \Psi (t)\rangle$ at its output. In the same basis, $| \Psi (t)\rangle$ is described by the column vector $ C_{+}(t)$ given by (\ref{C+(tzero)}), with $t$ replacing $t_{0}.$ In the matrix formalism, the relation between $C_{+}(t_{0})$ and $C_{+}(t)$ is written as
\begin{equation}
C_{+}(t)=MC_{+}(t_{0}),
\end{equation}%
where the square fourth-order matrix $M$ describes the effect of the coupling. In Ref. \cite{Deville2012} it was shown that when the coupling may be described by a Heisenberg cylindrical Hamiltonian, then $M=QDQ^{-1},$ where $Q=Q^{-1}$ is a square matrix with the following non-zero matrix elements:
\begin{equation}
Q_{11}=Q_{44}=1,\quad Q_{22}=-Q_{33}=Q_{23}=Q_{32}=\frac{1}{\sqrt{2}}
\end{equation}
and $D$ is a Diagonal square matrix with its diagonal elements equal to $D_{ii}=e^{-i\omega _{i}(t-t_{0})}$ ($i=1$...$4$), the $\omega _{i}$ being real quantities depending upon $J_{z}$ and $J_{xy}$, with generally unknown numerical values. The input of the inverting block then receives this state $|\Psi (t)\rangle.$ Its output provides a state $|\Phi \rangle$ described in the $\mathcal{B}_{+}$ basis by a column vector $C,$ with
\begin{equation}
C=UC_{+}(t)=UMC_{+}(t_{0}),  \label{C=UMC+(Tzero)}
\end{equation}%
where the square matrix $\ U$ (Unmixing matrix) describes the effect of the
inverting block of the separating system. If it\ is possible to choose $U$
in the form $U=M^{-1}$, then $| \Phi \rangle$ will be equal to $| \Psi (t_{0})\rangle.$\
But, strictly speaking, operating this way is impossible, because $M=QDQ,$ and $D$ is
unknown. In Ref. \cite{Deville2013} the inverting block was formally built using
a chain of quantum gates globally realizing matrix $U$ in the form $U=Q%
\widetilde{D}Q,$ where $\widetilde{D}$ 
is a diagonal matrix with its four diagonal elements $\widetilde{D}_{ii}$ ($i$ $=1$ ...4)~equal to%
\begin{equation}
\widetilde{D}_{ii}=e^{i\gamma _{i}}, \qquad \gamma_{i}: \mathrm{free\; real \;parameters.}
\end{equation}%
$\widetilde{D}D=\Delta $ is therefore a diagonal matrix with diagonal elements $\Delta _{ii}=e^{i\delta _{i}},$ where
\begin{equation}
\delta _{i}=\gamma _{i}-\omega _{i}(t-t_{0}).
\end{equation}%
The $\widetilde{D}$ matrix and the adaptation phase were introduced because it is not possible to modify the values of the $D$ matrix. In the following discussion, it is assumed that the $\omega _{i}$ are time-independent and that the adaptation phase has been successful with respect to unentanglement, i.e. that it has been possible to adjust the $\gamma _{i}$ in such a way that, in the inversion phase, if the Writer has prepared each qubit of the qubit pair in an arbitrary pure state at time $t_{0}$, we are then sure that state $| \Phi \rangle$ at the output of the inverting block is unentangled. The column vectors $C_{+}(t_{0})$ and $C$ are associated with $| \Psi (t_{0})\rangle$ and $| \Phi\rangle$ respectively, and $C=Q\Delta QC_{+}(t_{0})$ is therefore the column vector
\begin{equation}
\left( 
\begin{array}{c}
e^{i\delta _{1}}c_{1}(t_{0}) \\ 
\lbrack e^{i\delta _{2}}(c_{2}(t_{0})+c_{3}(t_{0}))+e^{i\delta
_{3}}(c_{2}(t_{0})-c_{3}(t_{0}))]/2 \\ 
\lbrack e^{i\delta _{2}}(c_{2}(t_{0})+c_{3}(t_{0}))-e^{i\delta
_{3}}(c_{2}(t_{0})-c_{3}(t_{0}))]/2 \\ 
e^{i\delta _{4}}c_{4}(t_{0})%
\end{array}%
\right) .  \label{VecteurColonneC_Phi}
\end{equation}
State $| \Phi \rangle$ is unentangled if and only if Eq. (\ref{c1c4=c2c3}) is
fulfilled, i.e. if
\begin{equation}
e^{i(\delta _{1}+\delta _{4})}c_{1}c_{4}=\frac{1}{4}%
[2c_{2}c_{3}(e^{i2\delta _{2}}+e^{i2\delta
_{3}})+(c_{2}^{2}+c_{3}^{2}(e^{i2\delta _{2}}-e^{i2\delta
_{3}})] \label{EtatenSortieNonIntrique1}
\end{equation}%
($c_{i}$ meaning $c_{i}(t_{0})$, for $i=1$ to $4$). We want this relation to be 
satisfied for any unentangled $|\Psi (t_{0})\rangle$. Starting with a $| \Psi (t_{0})\rangle$ 
state with $c_{2}(t_{0})c_{3}(t_{0})\neq 0$ and\ remembering that $
c_{1}(t_{0})c_{4}(t_{0})=c_{2}(t_{0})c_{3}(t_{0})$, Eq. (\ref{EtatenSortieNonIntrique1})
may then be written
\begin{equation}
e^{i(\delta _{1}+\delta _{4})}-\frac{1}{2}(e^{i2\delta _{2}}+e^{i2\delta
_{3}})=\frac{c_{2}^{2}(t_{0})+c_{3}^{2}(t_{0})}{4c_{2}(t_{0})c_{3}(t_{0})}%
(e^{i2\delta _{2}}-e^{i2\delta _{3}}).  \label{EtatenSortieNonIntrique2}
\end{equation}
Eq. (\ref{EtatenSortieNonIntrique2})\ is required to be fulfilled for all
possible states $| \Psi (t_{0})\rangle$ with $c_{2}(t_{0})c_{3}(t_{0})$ $\neq 0,$
and for fixed $\delta _{i}$ values (defined once for all during the
adaptation phase). Since the left-hand term does not depend upon the $c_{i}(t_{0})$
whereas its right-hand term does depend upon them, Eq. (\ref%
{EtatenSortieNonIntrique2}) is satisfied only if%
\begin{equation}
e^{i2\delta _{2}}-e^{i2\delta _{3}}=0,\quad \mathrm{i.e.}\quad \delta_{3}-\delta _{2}=m\pi ,\; m:
\mathrm{integer,} \label{NonIntrication1Delta1-4}
\end{equation}%
and then Eq. (\ref{EtatenSortieNonIntrique2}) moreover imposes that
\begin{equation}
\delta_{1}+\delta _{4}=2\delta _{2}+2k\pi ,\qquad k: \mathrm{integer.}
\label{NonIntrication1Delta2-3}
\end{equation}
If Eqs. (\ref{NonIntrication1Delta1-4}) and (\ref{NonIntrication1Delta2-3}) and relation $c_{1}(t_{0})c_{4}(t_{0})=c_{2}(t_{0})c_{3}(t_{0})$ are inserted into Eq. (\ref{VecteurColonneC_Phi}), it is easy to write $| \Phi \rangle$ as a
product state, which confirms that if Eq. (\ref{NonIntrication1Delta1-4}) and Eq.(\ref{NonIntrication1Delta2-3}) are fulfilled, then $|\Phi \rangle$ is unentangled indeed.

If one now supposes e.g. a $| \Psi(t_{0})\rangle$ with $c_{3}(t_{0})=0$, $ c_{2}(t_{0})\neq 0,$ $c_{4}(t_{0})\neq 0,$ and therefore $c_{1}(t_{0})=0$, then in order for $|\Phi  \rangle$ to be unentangled Eq. (\ref{EtatenSortieNonIntrique1}) has to be fulfilled. Putting $c_{1}(t_{0})=c_{3}(t_{0})=0$ in Eq. (\ref{EtatenSortieNonIntrique1}) leads to Eq. (\ref{NonIntrication1Delta1-4}), and the $\delta _{i}$ are then not submitted to another constraint. The same behavior is found if $c_{4}(t_{0})=c_{3}(t_{0})=0,$ and $c_{1}(t_{0})\neq 0,$ $c_{2}(t_{0})\neq 0$, and this remains true if $c_{1}(t_{0})=$ $c_{2}(t_{0})=c_{4}(t_{0})=0,$ $c_{3}(t_{0})\neq 0.$

When one starts with an \textit{arbitrary} initial unentangled state $| \Psi (t_{0})\rangle,$ the following property is a consequence of the results of the previous discussion. If during the adaptation phase it has been possible to rightly fix the $\gamma _{i}$ values, one may claim that the corresponding $|\Phi\rangle$ is unentangled if and only if during that adaptation phase the choice of the $\gamma _{i}$ has allowed conditions (\ref%
{NonIntrication1Delta1-4}) and (\ref{NonIntrication1Delta2-3}) to be \textit{%
both }fulfilled. This however does not guarantee that $|\Phi\rangle$ is identical to $|\Psi (t_{0}) \rangle$. The latter identification corresponds to source restoration itself, outside the scope of this article.
\section{Conclusion}
When trying to extend Blind Source Separation (BSS) from the Classical to the Quantum Information domain, with qubits realized with spins 1/2, one has to face two major consequences of the quantum context. First, if each qubit of a spin qubit pair is initially prepared in a pure state, and the time evolution of the pair state is governed by some undesired coupling between the spins, the Reader at the Mixer output accesses an unknown generally entangled qubit pair quantum state. This entangled state may be sent to a quantum processing system in order to restore the initially prepared state. Writing the output state of this processing system as $| \Phi  \rangle=\sum_{i}c_{i}\mid i \rangle$, in the standard basis, with well-ordered basis states, we showed that this state is unentangled if and only if $c_{1}c_{4}=c_{2}c_{3}$, a constraint between probability amplitudes. And secondly, results of measurements of the qubit spin components have a probabilistic nature, and the corresponding probabilities follow quantum properties even when processed with classical means. This article shows precautions to be taken when trying to extend to Blind Quantum SS the concept of source statistical independence used in conventional BSS. Using the probabilities $P_{izj}$ of getting the different possible results when measuring $s_{1z}$ and $s_{2j},$ successively with $j=z,$ $x$ and $y$, it is shown that the above unentanglement criterion may be written as $\{P_{1zj}P_{4zj}=P_{2zj}P_{3zj}\}$, a set of three constraints between probabilities. This unentanglement criterion could be of help in the adaptation phase of Blind Quantum SS, through some disentanglement-based separation principle, before restoration of the initial unentangled state.


\end{document}